\documentclass[paper,notoc,nohyper]{JHEP}
\usepackage{epsfig}
\usepackage{amsbsy} 
\def\lib{{\rm Li}_2}

\def\CA{C_A}
\def\CF{C_F}

\renewcommand\O[1]{{\cal O}\left(#1\right)}

\def\a0{\alpha_0}

\def\({\left(}
\def\){\right)}

\def\ket#1{|{#1}\rangle}

\def\cm{{\cal M}}

\def\bom#1{{\mbox{\boldmath $#1$}}}

\def \ep{\epsilon}

\title{\boldmath Second order contributions to elastic large-angle Bhabha
scattering
\footnote{Work supported in part by the UK Particle Physics and Astronomy 
Research Council, by the EU Fifth Framework Programme `Improving Human
Potential', Research Training Network `Particle Physics Phenomenology 
at High Energy Colliders', contract HPRN-CT-2000-00149 and
by the DFG-Forschergruppe
Quantenfeldtheorie, Computeralgebra und Monte-Carlo Simulation.
We thank the
British Council and German Academic Exchange Service for support under ARC
project 1050.} } 
\author{ E.~W.~N.~Glover$^a$, J.~B.~Tausk$^b$ and J.~J.~van der Bij$^b$\\ 
$^a$Department of 
Physics, University of Durham,  Durham DH1 3LE,  England\\[1mm]
$^b$Fakult{\"a}t f{\"u}r Physik,
        Albert-Ludwigs-Universit{\"a}t Freiburg,
        D-79104 Freiburg, Germany\\[1mm] 
E-mail:             \email{E.W.N.Glover@durham.ac.uk},
\email{tausk@physik.uni-freiburg.de}, \email{jochum@physik.uni-freiburg.de}} 
\abstract{ 
We derive the coefficient of the $\O{\alpha^2\log\left(s/m_e^2\right)}$
fixed order contribution to elastic large-angle Bhabha scattering.
We
adapt the classification of infrared divergences, that was recently
developed within dimensional regularization, and apply it  to
the regularization scheme with a massive photon and electron.}

\preprint{{DCTP/01/52}, {IPPP/01/26}, {Freiburg-THEP 01/05}, {hep-ph/0106052}}

\begin{document}

\section{Introduction}

Bhabha scattering has been measured with great accuracy by the experiments
at the LEP and SLC colliders. This has led to many works by numerous authors
being devoted to calculating the theoretical cross section for this
process as accurately as possible~\cite{Montagna,jadachnicrosini}.
Yet, the current accuracy of theoretical predictions for large angle
Bhabha scattering at LEP2 is still not completely satisfactory~\cite{kobel},
and limits the bounds that can be set on some kinds of potential new
physics effects, for instance, those coming from large extra
dimensions~\cite{bourilkov}.

Future linear electron positron colliders, e.g. the proposed TESLA
accelerator, will not have monochromatic beams because the electrons
and positrons can emit beamstrahlung before they collide. Measurement
of the resulting acollinearity angle in large angle Bhabha events
provides a way to determine the luminosity spectrum~\cite{toomi,tesla}.

Large angle Bhabha scattering is also important at electron-positron
colliders running at centre of mass energies of a few GeV, such as BEPC,
VEPP-2M, DAPHNE, and the B-factories PEP-II and KEK-B, where it is
used to measure the integrated luminosity~\cite{carloni}. At present,
the theoretical uncertainty on the differential cross section of this
process is one of the limiting factors on the precision of the luminosity
determination~\cite{babarosaka}.

In order to improve the theoretical predictions, it is necessary to
include higher order radiative corrections. In this paper, we consider
only pure QED corrections. These fall in different orders of magnitude
$\alpha L$, $\alpha$, $\alpha^2L^2$, $\alpha^2L$, $\alpha^3L^3$, where
$\alpha$ is the fine-structure constant and $L= \log (s/m_e^2)$. The large
logarithm $L$ is related to collinear divergences that would appear if the
electron mass $m_e$ were zero. The $\O{\alpha^2 L}$ terms are so far
only partially known. Arbuzov, Kuraev and Shaikhatdenov have calculated
the contributions from soft one- and two-photon bremsstrahlung, squared
one-loop graphs, and the interference between two-loop vertex graphs
and tree level terms~\cite{arbuzov}. However, they did not calculate
the contribution due to two-loop box graphs, which has to be included
to complete the $\O{\alpha^2 L}$ correction.

In principle, it should be possible to extract the missing two-loop box
contribution from the work of Bern, Dixon and Ghinculov~\cite{BDG}, who
have presented a complete formula for the two-loop virtual corrections to
Bhabha scattering. However, these authors set the electron mass to
zero and regularized all infrared and collinear divergences dimensionally,
with $d=4-2 \epsilon$, unlike the authors of ref~\cite{arbuzov}, who
used the traditional method of an electron and a photon mass. For this
reason, the results obtained by the two groups cannot be directly combined.

If one is only interested in the logarithmically enhanced term,
one does not actually need the full result of ref.~\cite{BDG}, but
only the terms containing powers of $1/\epsilon$. The correspondence
between poles in $\epsilon$ of the dimensional regularization scheme
and logarithms $\log(m_\gamma)$ and $\log(m_e)$ of the massive
regulator scheme has been worked out in detail at the one-loop
level~\cite{catanidittmaiertrocsanyi}. However, it is not a priori clear
how to extend this to two loops. Therefore, in this paper, we take
a slightly different approach. We rely on an elegant classification,
proposed by Catani, of the singularities of on-shell two-loop amplitudes
in QCD~\cite{catani}. His results were confirmed for the QED case
by~\cite{BDG}. Catani's formalism was developed using dimensional
regularization. However, it ought to be valid in both regularization
schemes. We shall translate this formalism into the massive regulator
scheme, and then use it to predict the $\O{\alpha^2 L}$ contribution to
the cross section for large angle Bhabha scattering. We limit ourselves
to graphs without photon vacuum polarization insertions, since those
can be treated as a separate class~\cite{arbuzovVSpairs}.

Our result should be directly applicable to the energy range
of a few GeV, where non-QED effects, from graphs involving
$Z$-boson exchange, are still very small. There are several Monte
Carlo programs that were specifically designed for this energy
range. These include the programs BABAYAGA and LABSPV~\cite{carloni}
and LABSMC~\cite{labsmc}. (Monte Carlo programs written for LEP1 and
LEP2 are reviewed in refs.~\cite{jadachnicrosini,kobel}). The precision
of these programs depends on the energy and on the details of the cuts
applied. In the case of LABSMC, under typical conditions, the uncertainty
is estimated by the authors to be around 0.2\%~\cite{labsmc}, mainly due
to missing terms of order $\O{\alpha^2 L}$. The result obtained in this
paper will allow this uncertainty to be reduced.

\section{Infrared factorisation formulae}

A generic (renormalized) QED matrix element 
can be expanded as a series in $\alpha$ as follows,
\begin{equation}
\ket{\cm}
=\left(\frac{\alpha}{2\pi}\right)^n \left(
\ket{\cm^{(0)}}+\left(\frac{\alpha}{2\pi}\right)\ket{\cm^{(1)}}
+\left(\frac{\alpha}{2\pi}\right)^2
\ket{\cm^{(2)}} + \O{\alpha^3} \right)
\end{equation}
where $\ket{\cm^{(i)}}$ represents the $i$-loop contribution and where the
overall power $n$ may be half-integer. According to Catani~\cite{catani}, these
amplitudes obey the following factorization formulae,
\begin{eqnarray}
\label{eq:factor}
\ket{\cm^{(1)}} &=& {\bom I}^{(1)}\ket{\cm^{(0)}}
+ \ket{\cm^{(1),fin}}\nonumber \\
\ket{\cm^{(2)}} &=& {\bom I}^{(2)}\ket{\cm^{(0)}} 
+ {\bom I}^{(1)}\ket{\cm^{(1)}} + \ket{\cm^{(2),fin}}
\end{eqnarray}
where ${\bom I}^{(1)}$ and ${\bom I}^{(2)}$ are operators that depend on the
regularization scheme and contain all of the singularities of the infrared 
regulator. The remainders $\ket{\cm^{(1),fin}}$ and
$\ket{\cm^{(2),fin}}$ are finite.
In dimensional regularization, for QED processes with massless electrons,
\begin{equation}
\label{eq:I1}
{\bom I}^{(1)}(\ep) = \frac{1}{2}\frac{e^{\ep \gamma}}{\Gamma(1-\ep)}
\sum_{i=1}^n \sum_{i\neq j}^n e_ie_j  \( \frac{1}{\ep^2}+ \frac{3}{2\ep}\)
\left( \frac{\mu^2 e^{-i\lambda_{ij}\pi}}{2p_i.p_j}\right)^{\epsilon} \, ,
\end{equation}
where $\lambda_{ij} = -1$ if $i$ and $j$ are both incoming or
outgoing partons and $\lambda_{ij}= 0$ otherwise, and $e_i$ is the
electric charge (minus the electric charge) of an outgoing (incoming)
radiating particle with momentum $p_i$.
Eq.~(\ref{eq:I1}) is obtained from the QCD result~\cite{catani} with
the replacements $\CA \to 0$, $\CF \to 1$, $T_R \to 1$ and
${\bom T}_i\cdot{\bom T}_j \to e_i e_j$.

Because we exclude graphs with photon vacuum polarization
insertions, we neglect terms proportional to the $\beta$ function.
The operator ${\bom I}^{(2)}$ is then given by
\begin{eqnarray}
{\bom I}^{(2)} &=&
 -\frac{1}{2} {\bom I}^{(1)}{\bom I}^{(1)} + {\bom H}^{(2)} \, ,
\end{eqnarray}
where ${\bom H}^{(2)}$ contains only single poles and has the form
\begin{equation}
\label{eq:H2bomdimreg}
{\bom H}^{(2)} = \frac{1}{4\epsilon} \frac{e^{\ep \gamma}}{\Gamma(1-\ep)} 
\sum_{i=1}^n \sum_{i\neq j}^n e_ie_j
 \left( \frac{\mu^2 e^{-i\lambda_{ij}\pi}}{2p_i.p_j}\right)^{2\epsilon}
H^{(2)},
\end{equation}
where 
\begin{equation}
\label{eq:h2}
H^{(2)} =    \frac{3}{8}-3\zeta_2+6\zeta_3. 
\end{equation}
Ref.~\cite{catani} does not give a general formula for $H^{(2)}$,
and at present it is necessary to derive it from an explicit two-loop
calculation, such as the two-loop QCD computation of the electromagnetic
form factor of the quark~\cite{willy}.
We note that in Ref.~\cite{catani} the summation over radiating pairs
$i$ and $j$ is not explicitly included in the definition of ${\bom
H}^{(2)}$. However, it has been found to
reproduce correctly the results of explicit calculations in QED~\cite{BDG}
and QCD~\cite{us}. We also note that the terms of $\O{1}$ in
${\bom H}^{(2)}$ are presently a matter of choice and can be altered by
a redefinition of $\ket{\cm^{(2),fin}}$.

\section{The massive photon and massive electron regularisation scheme}

The factorization formulae~(\ref{eq:factor}) should hold for any choice of
infrared regulator. Therefore, we can find a translation between schemes by
comparing explicit calculations. For example, the one-loop correction
to the electron photon vertex (see, e.g.  Ref.~\cite{BMR}),
serves to fix ${\bom I}^{(1)}$ in the scheme where the photon has 
mass $\lambda$ and the electron has mass $m$.
We write the Dirac form factor $F_1(q^2)$ of the electron photon vertex
\begin{equation}
\Gamma_{\mu} =
\gamma_{\mu} F_1(q^2) + \frac{i}{2 m} \sigma_{\mu \nu} q^{\nu} F_2(q^2)
\end{equation}
as\footnote{
In order to follow more closely the notation of
Ref.~\cite{arbuzov}, we expand in $\alpha/{\pi}$ rather than
$\alpha/{(2\pi)}$ in the remainder of this paper.  
Therefore, the definitions of ${\bom I}^{(1)}$ and  ${\bom H}^{(2)}$ 
differ by factors of $1/2$ and  $1/4$ respectively from those in the previous
section.}
\begin{equation}
F_1(q^2) = 1 + \frac{\alpha}{\pi} \, F_1^{(1)} (q^2)
 + {\left(\frac{\alpha}{\pi}\right)}^2 F_1^{(2)} (q^2)
 + \O{\alpha^3} \, .
\end{equation}
By equating
\begin{equation}
 F_1^{(1)} (q^2) = {\bom I}^{(1)}(q^2) +  F_1^{(1),fin} (q^2) \, ,
\end{equation}
we see that
\begin{equation}
{\bom I}^{(1)}(q^2)=(L-1)\left(\frac{1}{2}L_{\lambda}+1\right)
-\frac{1}{4}L^2-\frac{1}{4}L+\frac{\pi^2}{12} \, ,
\end{equation}
where $L_{\lambda} = \log\left(\lambda^2/m^2\right)$,
$L = \log\left(-q^2/m^2\right)$ if $q^2 < 0$ and $L =
\log\left(q^2/m^2\right)-i\pi$ if $q^2 > 0$.
There is a possible ambiguity in assigning the constant pieces to 
${\bom I}^{(1)}$ or $F_1^{(1),fin}$.  Our choice corresponds to 
$F_1^{(1),fin}=0$.
Similarly, by examining the two-loop vertex, we can fix ${\bom H}^{(2)}$.
To do this, we take the large scale limit of the $\O{e^4}$ two-loop vertex
correction
$F_1^{(2)}(q^2)$ computed by
Barbieri, Mignaco and Remiddi~\cite{BMR} (without the contribution
from the vacuum polarization graph). Up to terms that do not depend on
the regulators, we expect that,
\begin{equation}
F_1^{(2)}(q^2) = \frac{1}{2}{\bom I}^{(1)}(q^2){\bom I}^{(1)}(q^2)
+ {\bom H}^{(2)}(q^2) + \O{1}.
\end{equation}
We find that ${\bom H}^{(2)}(q^2)$ is proportional to a single large logarithm 
\begin{equation}
{\bom H}^{(2)}(q^2) = \frac{1}{4}\, L H^{(2)}
\end{equation}
where $H^{(2)}$ is given by Eq.~(\ref{eq:h2}).
In fact, one might wonder whether changing the constants in
${\bom I}^{(1)}(q^2)$ would give rise to a different $H^{(2)}$. This is not
the case since any change is absorbed by the necessary alteration to
$F_1^{(1),fin}$.

Armed with these operators using mass regularization, we can compute the
large logarithmic corrections to the two-loop contribution to Bhabha
scattering keeping the electron mass and using a small photon mass as
the infrared regulator.
For $2 \to 2$ scattering, there are six radiating pairs, so that  
\begin{equation}
\label{eq:i1actual}
{\bom I}^{(1)}
= 2\left({\bom I}^{(1)}(s)+{\bom I}^{(1)}(t)-{\bom I}^{(1)}(u)\right),
\end{equation}
and
\begin{equation}
\label{eq:h2actual}
{\bom H}^{(2)}
= 2\left({\bom H}^{(2)}(s)+{\bom H}^{(2)}(t)-{\bom H}^{(2)}(u)\right).
\end{equation}

The second order virtual contribution to Bhabha scattering comes from
the square of the one-loop graphs and the interference of tree and
two-loop graphs. We can write this contribution in factorized form as
\begin{equation}
\frac{{\rm d} \sigma^{VV}}{{\rm d} \sigma_0} =
  \left(\frac{\alpha}{\pi}\right)^2 \Delta_{VV} \, ,
\end{equation}
where the lowest order differential cross section is given by
\begin{equation}
{\rm d}\sigma_0=\frac{\alpha^2}{s}
\left(\frac{1-x+x^2}{x}\right)^2 {\rm d}\Omega \, ,
\end{equation}
with $x = -t/s$ and $1-x = -u/s$.
Up to constant terms, 
\begin{equation}
\label{eq:dvv}
\Delta_{VV} = -\frac{1}{2} \left({\bom I}^{(1)}+{\bom I}^{(1)*}\right)^2
+ \delta_V  \left({\bom I}^{(1)}+{\bom I}^{(1)*}\right) +
\left({\bom H}^{(2)}+{\bom H}^{(2)*}\right)
\end{equation}
with ${\bom I}^{(1)}$ and ${\bom H}^{(2)}$ given by 
Eqs.~(\ref{eq:i1actual}) and (\ref{eq:h2actual}) respectively.
The second term involves the one-loop virtual contribution
$\delta_V$ (see, e.g.~\cite{BerendsKleiss,CaffoGattoRemiddi} and
references therein). In the notation of ref.~\cite{arbuzov},
it is given by,
\begin{eqnarray}
\label{eq:dv}
\delta_V&=&4\log\frac{m}{\lambda}
\left(1-L+\log\left(\frac{1-x}{x}\right)\right)
-L^2+2L\log\left(\frac{1-x}{x}\right)-\log^2(x) \nonumber \\ 
&+& \log^2(1-x) + 3L-4+f(x),
\end{eqnarray}
with\footnote{There are some misprints in the formula for
$f(x)$ in ref.~\cite{arbuzov}.}
\begin{eqnarray}
f(x)&=&(1-x+x^2)^{-2}\biggl[
\frac{\pi^2}{12}( 4 - 8 x + 27 x^2 - 26 x^3 + 16 x^4)\nonumber \\
&& +\frac{1}{2}(-2+5x-7x^2+5x^3-2x^4)\log^2(1-x)
+\frac{1}{4}x(3-x-3x^2+4x^3)\log^2(x)\nonumber \\
&&
+\frac{1}{2}(6 - 8 x + 9 x^2 - 3 x^3)\log(x)
-\frac{1}{2}x(1+x^2)\log(1-x)\nonumber \\
&&
+\frac{1}{2}(4-8x+7x^2-2x^3)\log(x)\log(1-x) \biggr].
\end{eqnarray}
Expanding $\Delta_{VV}$ yields all terms containing at least one power
of the large logarithm $L = \log(s/m^2)$ as well as all logarithms of
the photon mass regulator.
We have checked that in the small angle limit, $x\to 0$,
to the logarithmic accuracy we are working at,
Eq.~(\ref{eq:dvv}) reduces to the known result~\cite{permille}
\begin{equation}
\Delta_{VV} = 6\, {\left(F_1^{(1)}(t)\right)}^2 + 4\, F_1^{(2)}(t) \, ,
\end{equation}
which follows from a generalized eikonal representation~\cite{eikonal}
for the Bhabha scattering amplitude for small angles.

\section{The scattering cross-section at $\O{\alpha^2}$}

To make a physical prediction, the double virtual contribution must be
combined with the one-loop contribution with single soft emission and
the tree level double soft emission. The second order correction to
the one-loop virtual photon emission corrected cross section, due to the
emission of a single real soft photon having energy less than
$\Delta\varepsilon$, can be written down in the factorized form~\cite{arbuzov}
\begin{equation}
\frac{{\rm d} \sigma^{SV}}{{\rm d} \sigma_0}=\frac{\alpha}{\pi}\delta_S
\frac{\alpha}{\pi}\delta_V=\left(\frac{\alpha}{\pi}\right)^2\Delta_{SV},
\end{equation}
where $\delta_V$ is given in Eq.~(\ref{eq:dv}) and
\begin{eqnarray}
\delta_S&=&4\log\left(\frac{m\Delta\varepsilon}{\lambda\varepsilon}\right)
\left(L-1+\log\left(\frac{x}{1-x}\right)\right)
+L^2+2L\log\left(\frac{x}{1-x}\right)+\log^2(x) \nonumber \\ 
&-&\log^2(1-x)-\frac{2\pi^2}{3}+2\,\lib(1-x)-2\,\lib(x).
\end{eqnarray}
Here, $\varepsilon=\sqrt{s/4}$ is the energy of the electron and positron
beams. The dilogarithm function is defined as
\begin{equation}
\lib(x) = - \int\limits_{0}^{x}\frac{{\rm d} y}{y}\log(1-y).
\end{equation}

The contribution from two independently emitted soft photons each
with energy $\omega_1,\omega_2\leq\Delta\varepsilon$ is given
by~\cite{arbuzov}
\begin{equation} 
\label{SS}
\frac{{\rm d} \sigma^{SS}}{{\rm d} \sigma_0}=
\frac{1}{2}\left(\frac{\alpha}{\pi} \right)^2\delta_S^2
 \equiv \left(\frac{\alpha}{\pi}\right)^2\Delta_{SS},
\end{equation}
where the statistical factor $1/2!$ is due to the identity of photons.

The combination $\Delta_{SS}+\Delta_{SV}$ is written in expanded form
in Ref.~\cite{arbuzov} where terms like $L^4$, $L^3 L_\lambda$, $L^3$,
$L^2 L_\lambda$, $L^2 L_\lambda^2$, $L L_\lambda$ and $L L_\lambda^2$ are
produced. All of these terms precisely cancel against similar terms in
$\Delta_{VV}$. The final second order contribution to the cross section
through to $\O{1}$ is given by
\begin{eqnarray}
\frac{{\rm d} \sigma}{{\rm d} \sigma_0} &=& 
 \left(\frac{\alpha}{\pi}\right)^2 \left(\Delta_{VV}+\Delta_{SV}+ \Delta_{SS}
 \right)\nonumber \\
 &=& \left(\frac{\alpha}{\pi}\right)^2 \left[ L^2 \left(
  8\log^2\left(\frac{\Delta\varepsilon}{\varepsilon}\right)
 +12 \log\left(\frac{\Delta\varepsilon}{\varepsilon}\right)
 +\frac{9}{2} \right) 
 \right. \nonumber \\
&&\phantom{\left(\frac{\alpha}{\pi}\right)^2}
 + \left. L \left(
 A \log^2\left(\frac{\Delta\varepsilon}{\varepsilon}\right)
+B \log\left(\frac{\Delta\varepsilon}{\varepsilon}\right)  
+C
\right)\right].
\label{eq:result}
\end{eqnarray}
The single logarithmic coefficients $A$, $B$ and $C$ are given by,
\begin{eqnarray}
A&=& 16 \log\left(\frac{x}{1-x}\right) -16,\\
B&=& 8\,\lib{(1-x)}-8\,\lib{(x)}+12\log\left(\frac{x}{1-x}\right) +4 f(x) -28
-\frac{8}{3}\pi^2,\\
C&=& 6\,\lib{(1-x)}-6\,\lib{(x)} + 3 f(x)
 + 6 \zeta_3 -\frac{93}{8} - \frac{5}{2} \pi^2.
\end{eqnarray}
Coefficients $A$ and $B$ agree with those obtained by
Arbuzov et al.~\cite{arbuzov}\footnote{up to a slight misprint
in the expression for $B$, denoted there by $z_1$},
while $C$ is the main new result of the present work.

\section{Summary}

In this form Eq.~(\ref{eq:result}) cannot be used to compare directly with
experiment. The reason is that an experimental set-up involves a
complicated detector that is not represented by the simple energy cuts
on the photons which we used here. Such effects have to be modeled by
a Monte-Carlo calculation. However any uncertainty in the cross section
is now reduced to $\O{\alpha^2}$ terms without enhanced logarithms. Given
the final simplicity of the method one can wonder whether it is possible
to use the results of~\cite{BDG} to extract the finite parts of the cross
section. At present this appears to be not simply possible, as we have
no way to tell whether the $1/\epsilon$ poles contained in
${\bom H}^{(2)}$ (Eq.~(\ref{eq:H2bomdimreg}))
correspond to $L$ or for instance $L-1$.
Presumably it is possible to establish this connection by further
expanding the vertex functions. The alternative would be to calculate
the box graphs within the mass regulator scheme. However this is quite
a challenge. For application to analogous processes in QCD we remark
however that this last uncertainty plays no role as it gets absorbed in
the definition of the structure functions.

The calculation presented in this paper removes a major obstacle
to improvement of the precision of Monte Carlo programs for
large angle Bhabha scattering at centre of mass energies of a few
GeV~\cite{labsmc}. The result can also be used at higher energies,
provided that additional contributions due to $Z$-exchange diagrams
are included. Whether it is possible to obtain them by an extension
of the method we have used here is a question that requires
further investigation.

\section*{Acknowledgements}

We thank Wim Beenakker for discussions.
This work was supported in part by the UK Particle Physics and Astronomy 
Research Council,  by the EU Fifth Framework Programme `Improving Human
Potential', Research Training Network `Particle Physics Phenomenology 
at High Energy Colliders', contract HPRN-CT-2000-00149 and
by the DFG-Forschergruppe
Quantenfeldtheorie, Computeralgebra und Monte-Carlo Simulation.
We thank the British Council and German Academic Exchange Service for
support under ARC project 1050.


\begin{thebibliography}{99}

\bibitem{Montagna}
G.~Montagna, O.~Nicrosini and F.~Piccinini,
%``Precision physics at LEP,''
Riv.\ Nuovo Cim.\  {\bf 21N9} (1998) 1
[hep-ph/9802302].
%%CITATION = HEP-PH 9802302;%%

\bibitem{jadachnicrosini}
S.~Jadach {\it et al.} in:
G.~Altarelli, T.~Sj{\"o}strand and F.~Zwirner (eds.),
``Physics at LEP2,''
CERN-96-01
%``Event Generators for Bhabha Scattering,''
[hep-ph/9602393].
%%CITATION = HEP-PH 9602393;%%

\bibitem{kobel}
M.~Kobel {\it et al.}
%``Two-fermion production in electron positron collisions,''
hep-ph/0007180.
%%CITATION = HEP-PH 0007180;%%

\bibitem{bourilkov}
D.~Bourilkov,
%``Global analysis of Bhabha scattering at LEP2 and limits on low scale
% gravity models,''
JHEP {\bf 9908} (1999) 006
[hep-ph/9907380];
%%CITATION = HEP-PH 9907380;%%
M.~Beccaria, F.~M.~Renard, S.~Spagnolo and C.~Verzegnassi,
%``Z-peak subtracted representation of Bhabha scattering and search for
% new physics effects,''
Phys.\ Rev.\ D {\bf 62} (2000) 053003
[hep-ph/0002101].
%%CITATION = HEP-PH 0002101;%% 

\bibitem{toomi}
N.~Toomi, J.~Fujimoto, S.~Kawabata, Y.~Kurihara and T.~Watanabe,
%``Luminosity spectrum measurement in future e+ e- linear colliders using
% large-angle Bhabha events,''
Phys.\ Lett.\ B {\bf 429} (1998) 162.
%%CITATION = PHLTA,B429,162;%%

\bibitem{tesla}
R.~D.~Heuer, D.~Miller, F.~Richard and P.~M.~Zerwas (eds.),
``TESLA Technical design report. Pt. 3: Physics at an e+ e- linear collider,''
DESY-01-011C.

\bibitem{carloni}
C.~M.~Carloni Calame, C.~Lunardini, G.~Montagna, O.~Nicrosini and F.~Piccinini,
%``Large-angle Bhabha scattering and luminosity at flavour factories,''
Nucl.\ Phys.\ B {\bf 584} (2000) 459
[hep-ph/0003268].
%%CITATION = HEP-PH 0003268;%%

\bibitem{babarosaka}
B.~Aubert {\it et al.}  [BABAR Collaboration],
%``The first year of the BaBar experiment at PEP-II,''
hep-ex/0012042.
%%CITATION = HEP-EX 0012042;%%

\bibitem{arbuzov}
A.B. Arbuzov, E.A. Kuraev and B.G. Shaikhatdenov,
Mod. Phys. Lett. {\bf A13} (1998) 2305 [hep-ph/9806215]. 

\bibitem{BDG}
Z. Bern, L. Dixon and A. Ghinculov,  Phys. Rev. {\bf D63}
(2001) 053007  [hep-ph/0010075].

\bibitem{catanidittmaiertrocsanyi}
S.~Catani, S.~Dittmaier and Z.~Trocsanyi,
%``One-loop singular behaviour of QCD and SUSY QCD amplitudes with
%  massive partons,''
Phys.\ Lett.\ B {\bf 500} (2001) 149
[hep-ph/0011222].
%%CITATION = HEP-PH 0011222;%%

\bibitem{catani}
S. Catani, Phys. Lett. {\bf B427} (1998) 161 [hep-ph/9802439].

\bibitem{arbuzovVSpairs}
A.~B.~Arbuzov, E.~A.~Kuraev, N.~P.~Merenkov and L.~Trentadue,
%``Virtual and soft real pair production in large angle Bhabha scattering,''
Phys.\ Atom.\ Nucl.\ {\bf 60} (1997) 591.
%%CITATION = PANUE,60,591;%%

\bibitem{labsmc}
A.~B.~Arbuzov, G.~V.~Fedotovich, E.~A.~Kuraev, N.~P.~Merenkov, V.~D.~Rushai and
L.~Trentadue,
%``Large angle QED processes at e+ e- colliders at energies below 3-GeV,''
JHEP {\bf 9710} (1997) 001
[hep-ph/9702262];
%%CITATION = HEP-PH 9702262;%%
A.~B.~Arbuzov,
%``LABSMC: Monte Carlo event generator for large-angle Bhabha scattering,''
hep-ph/9907298.
%%CITATION = HEP-PH 9907298;%%

\bibitem{willy}
R.~J.~Gonsalves,
%``Dimensionally Regularized Two Loop On-Shell Quark Form-Factor,''
Phys.\ Rev.\ D {\bf 28} (1983) 1542;
%%CITATION = PHRVA,D28,1542;%%
G.~Kramer and B.~Lampe,
%``Two Jet Cross-Section In E+ E- Annihilation,''
Z.\ Phys.\ C {\bf 34} (1987) 497
[Erratum-ibid.\ C {\bf 42} (1987) 504];
%%CITATION = ZEPYA,C34,497;%%
T.~Matsuura and W.~L.~van Neerven,
%``Second Order Logarithmic Corrections To The Drell-Yan Cross-Section,''
Z.\ Phys.\ C {\bf 38} (1988) 623;
%%CITATION = ZEPYA,C38,623;%%
T.~Matsuura, S.~C.~van der Marck and W.~L.~van Neerven,
%``The Calculation Of The Second Order Soft And Virtual
% Contributions To The Drell-Yan Cross-Section,''
Nucl.\ Phys.\ B {\bf 319} (1989) 570.
%%CITATION = NUPHA,B319,570;%%

\bibitem{us}
C.~Anastasiou, E.~W.~N.~Glover, C.~Oleari and M.~E.~Tejeda-Yeomans,
%``Two-loop QCD corrections to q anti-q $\to$ q' anti-q',''
Nucl.\ Phys.\ B {\bf 601} (2001) 318 [hep-ph/0010212];
%%CITATION = HEP-PH 0010212;%%
%``Two-loop QCD corrections to q anti-q $\to$ q anti-q,''
Nucl.\ Phys.\ B {\bf 601} (2001) 341 [hep-ph/0011094];
%%CITATION = HEP-PH 0011094;%%
%``Two-loop QCD corrections to massless quark gluon scattering,''
hep-ph/0101304;
%%CITATION = HEP-PH 0101304;%%
E.~W.~N.~Glover, C.~Oleari and M.~E.~Tejeda-Yeomans,
%``Two-loop QCD corrections to gluon gluon scattering,''
hep-ph/0102201.
%%CITATION = HEP-PH 0102201;%%

\bibitem{BMR}
R. Barbieri, J. Mignaco and E. Remiddi, Il Nuovo Cimento {\bf A11}
(1972) 824; Il Nuovo Cimento {\bf A11} (1972) 865.

\bibitem{BerendsKleiss}
F.~A.~Berends and R.~Kleiss,
%``Distributions In The Process E+ E- $\to$ E+ E- (Gamma),''
Nucl.\ Phys.\ B {\bf 228} (1983) 537.
%%CITATION = NUPHA,B228,537;%%

\bibitem{CaffoGattoRemiddi}
M.~Caffo, R.~Gatto and E.~Remiddi,
%``Hard Collinear Photons. High-Energy Radiative Corrections
%To Bhabha Scattering,''
Nucl.\ Phys.\ B {\bf 252} (1985) 378.
%%CITATION = NUPHA,B252,378;%%

\bibitem{permille}
A.~B.~Arbuzov, V.~S.~Fadin, E.~A.~Kuraev, L.~N.~Lipatov, N.~P.~Merenkov
and L.~Trentadue,
%``Small-angle electron positron scattering with a per mille accuracy,''
Nucl.\ Phys.\ B {\bf 485} (1997) 457
[hep-ph/9512344].
%%CITATION = HEP-PH 9512344;%%

\bibitem{eikonal}
V.~S.~Fadin, E.~A.~Kuraev, L.~Trentadue, L.~N.~Lipatov and N.~P.~Merenkov,
%``Generalized eikonal representation of the small angle
% e+ e- scattering amplitude at high-energy,''
Phys.\ Atom.\ Nucl.\  {\bf 56} (1993) 1537
[Yad.\ Fiz.\  {\bf 56N11} (1993) 145].
%%CITATION = PANUE,56,1537;%%

\end{thebibliography}
\end{document}